# Will Dynamic Arrays finally change the way Models are built?

Peter Bartholomew
MDAO Technologies Ltd
peter.bartholomew1@btconnect.com

**ABSTRACT**

*Spreadsheets offer a supremely successful and intuitive means of processing and exchanging numerical content. Its intuitive ad-hoc nature makes it hugely popular for use in diverse areas including business and engineering, yet these very same characteristics make it extraordinarily error-prone; many would question whether it is suitable for serious analysis or modelling tasks.*

*A previous EuSpRIG paper examined the role of Names in increasing solution transparency and providing a readable notation to forge links with the problem domain. Extensive use was made of CSE array formulas, but it is acknowledged that their use makes spreadsheet development a distinctly cumbersome task. Since that time, the new dynamic arrays have been introduced and array calculation is now the default mode of operation for Excel. This paper examines the thesis that their adoption within a more professional development environment could replace traditional techniques where solution integrity is important.*

*A major advantage of fully dynamic models is that they require less manual intervention to keep them updated and so have the potential to reduce the attendant errors and risk.*

## 1 INTRODUCTION

This paper starts by reviewing the decisions made at the time electronic spreadsheet was invented and looks at the impact of those original decisions. Dan Bricklin required a means for recording the parameters used in a formula which is both usable for the calculation and intelligible for the user. Dan was well aware of the "programmer's way" of achieving this through the use of named variables but instead plumped for a strategy that was more action-led and intuitive. Thus, the co-ordinate approach of identifying data by location on a grid was created. This required no effort on behalf of the use to define either the terms they wished to use or their relationship to business objects; to do so was deemed 'tedious'.

These early decisions still characterise spreadsheet use today. Catering specifically for 'end user computing' has allowed the spreadsheet to become the most widely used and understood program throughout the world. It is the most widely used medium for the exchange of numerical data and across various business areas domain-specific software is likely to feature an 'Export to Excel' button.

Although professionally developed spreadsheets tend to be less haphazard in their construction, it is something of a surprise that the same primitive spreadsheet methods still dominate most spreadsheet development. The consistent use of defined names, array formula, structured references, pivot tables, Power Query and Power Pivot is the exception rather than the rule.

In place of adopting strategies that impose greater structure through the coding techniques employed, some of the financial modelling standards approach the problem by defining rules and standard practices to constrain the user to somewhat more ordered practices. This may start to address the 'Wild West' symptoms of endemic spreadsheet risk but does little to address the underlying disease.

## 2 THE SPREADSHEET CHALLENGE

### 2.1 Applicability

Standard Excel practice merely captures what the user does but provides no real indication of their intent. Worse, it does it in the most obscure manner imaginable *e.g.* 'take the number one cells above wherever it is that your formula happens to reside and add to it the number you find 6 cells to the right'. What on earth does the position of two numbers on a worksheet tell you about the business problem? Absolutely nothing; such an approach utterly fails to capture any business logic or the user-intent.

This bizarre process is then usually followed by 'and then copy the formula down': why, one might ask, how far? The action implies the existence of an unacknowledged list or array. Any such structure that may exist with the problem domain has to be inferred from the pattern of formula references that results from user actions (bearing in mind that each action carries with it something like a 2% chance of being an error, Panko (2015).

Almost always, the business objects of interest turn out not to be single cells; they may be arrays representing a quantity varying over time or lists of related objects such as that contained within a database table. This is evident in the case of financial models where one of the few things that is universally accepted is that all formulas should be uniform across the time period of the model. Such uniformity is the defining property of an array formula, yet the standards tend to either reject the use of array formulas or at least regard them as methods of last resort. Only if the syntax captures these objects and relates them to domain knowledge does one have anything of value. Without the ability to name objects at an appropriate level of abstraction, one must resort to annotation as a means of communicating the significance of content. One might hope to see the annotation in reasonable proximity to the range it describes but there is no real link between such labels and the data.

It should be noted, that although such criticism of the concept of direct cell referencing, as well as the merging of presentation, formulas and values as single cell objects is unusual in the world of spreadsheets, others have raised similar objections over the years, Hellman (2001), and the companies behind Analytica, Quantrix and related software have also had harsh criticisms, even ridicule, of spreadsheet technology. Finally, a quote from the Quantrix website:

> "*Traditional spreadsheets use cell addresses – such as B2 for column B of row 2 – rather than meaningful names – such as Revenues or Expenses. Anyone who has tried to read such formulas, whether written by someone else or oneself, knows how hard they can be to understand or verify*".

Not many such critics still use Excel though.

## 3 THIS IS THE WAY IT HAS BEEN SINCE THE BEGINNING; WHY CHANGE NOW?

Last September fundamental changes to the way in which Excel both stores data and performs calculation were announced at the Ignite Conference held in Florida. Prior to that, the value property of a single cell was limited to a single value, be it a String, a Double, a Boolean or an Error. At the Florida meeting Joe McDaid of Microsoft described how the newest releases of Excel could associate array object with a single cell.

Before, the elements of an array would have to be individually stored in a contiguous range of cells, and formulas would be implemented by the user on a cell-by-cell basis. Clearly, this would be horribly inconvenient if the references within each formula had to be built individually but, unless one of the referenced arrays happens to be transposed, the relative position of the formula cell and the cells it references will be uniform throughout the calculation. It is this property that was exploited with the introduction of relative referencing as a spreadsheet default; a relative reference has no intrinsic meaning, but it does serve to associate the correct values across the remaining cells of a referenced array. References in the form RC[-2] make the relationship explicit, whilst B7 conceals the relative nature of the reference behind a notation that at first sight appears to be absolute. It is the use of this 'hybrid' notation that gives rise to the characteristic pattern of references seen on spreadsheets.

In the new Excel all that changes. Any formula that references one or more arrays will return an array; it is no longer necessary to reference cells one by one using relative references to build an array result. The formula is contained within a single cell, though adjacent empty cells are needed in order to display the result, a process described as 'spilling'. The notation used for the resulting array is the cell reference followed by the '#' character. The reference to the anchor cell may be a direct reference or a defined name, though in published work to date the former is more commonly seen

## 4 ARRAY FORMULAS IN MODERN EXCEL

Array formulas create amazingly simple solutions. Instead of a multitude of formulas that should be the same but are in no way constrained to be so, the formula resides in a single cell. The thing that has changed is that a cell can now accept a complete array as its 'value' property whereas previously it was limited to taking a scalar value, be it a string, a number, a Boolean or an error.

Clearly, it is not possible to display an array within a single cell, so what is done instead is to use adjacent blank cells to show individual values, a process referred to as 'spilling'. This will be demonstrated in the example section below.

If one wishes to link the formulas back to the business problem, it is only these single cells that need to be named. If a name such as 'revenue' is applied to the anchor cell containing the formula then the entire array can be referenced as 'revenue#' and, as such the reference will adjust dynamically to match the content.

Whilst the headline functionality of the new modern Array Formulas is to make array calculation the default mode of operation, the new functions that have accompanied the release have a strong focus upon data analysis and list operations. Here, the distinction I am making is that an array is ordered, so allowing elements to be returned by index, whereas the significance of a list does not depend upon order. An array often represents a function of some independent variable, time being the most important example, but many other possibilities exist such as random variables within a probability calculation. The key difference is that operations such as filtering and sorting are not valid operations applied to an array whereas they are a mainstay of managing lists; a calendar sorted alphabetically is not much use to anyone!

Despite the distinction made here, there is a huge overlap that permits lists to be treated as arrays and conversely. In the following section, I will demonstrate the dynamic array formulas used to create a crosstab report. Of course, for large datasets, one would use PowerPivot but for small to medium datasets the new array functions offer a viable and very responsive alternative.

## 5 EXAMPLE – DATA ANALYSIS

The following example is based upon a hypothetical computer business with nation-wide sales. The data table has 5000 rows and the objective is to create a crosstab summary using formulas.

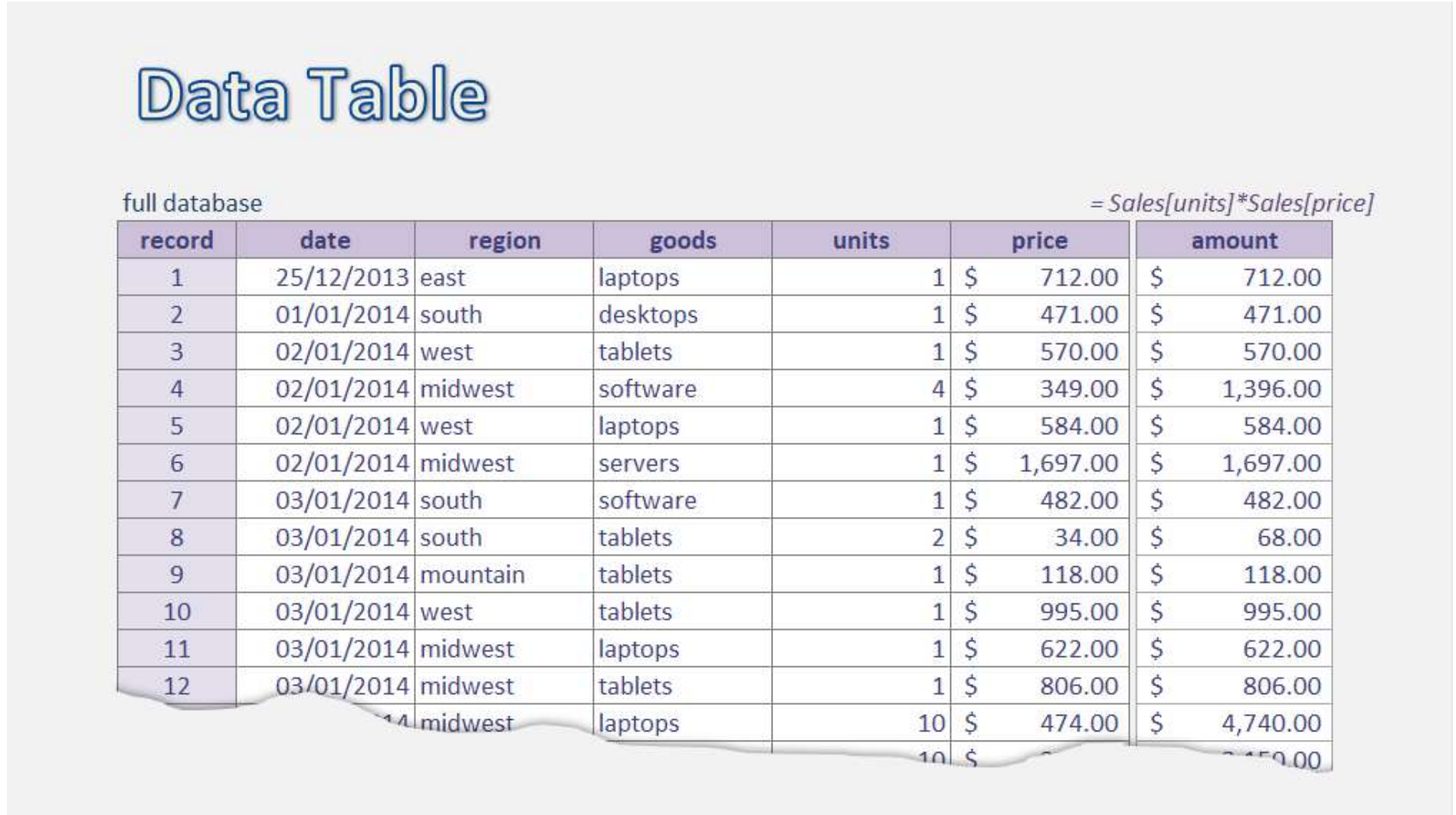

Data Table

full database

= Sales[units]*Sales[price]

| record | date | region | goods | units | price | amount |
|---|---|---|---|---|---|---|
| 1 | 25/12/2013 | east | laptops | 1 | $ 712.00 | $ 712.00 |
| 2 | 01/01/2014 | south | desktops | 1 | $ 471.00 | $ 471.00 |
| 3 | 02/01/2014 | west | tablets | 1 | $ 570.00 | $ 570.00 |
| 4 | 02/01/2014 | midwest | software | 4 | $ 349.00 | $ 1,396.00 |
| 5 | 02/01/2014 | west | laptops | 1 | $ 584.00 | $ 584.00 |
| 6 | 02/01/2014 | midwest | servers | 1 | $ 1,697.00 | $ 1,697.00 |
| 7 | 03/01/2014 | south | software | 1 | $ 482.00 | $ 482.00 |
| 8 | 03/01/2014 | south | tablets | 2 | $ 34.00 | $ 68.00 |
| 9 | 03/01/2014 | mountain | tablets | 1 | $ 118.00 | $ 118.00 |
| 10 | 03/01/2014 | west | tablets | 1 | $ 995.00 | $ 995.00 |
| 11 | 03/01/2014 | midwest | laptops | 1 | $ 622.00 | $ 622.00 |
| 12 | 03/01/2014 | midwest | tablets | 1 | $ 806.00 | $ 806.00 |
| | | midwest | laptops | 10 | $ 474.00 | $ 4,740.00 |

**Figure 1 Extract taken from source data table**

The table itself is conventional with values providing information relating to sales volume and unit price, combined with other fields that provide further information related to the transaction. Excel tables support calculated fields and the 'amount' could well have been included as part of the table. Since the aim here is to discuss the potential impact of modern dynamic arrays an alternative route is taken in this case.

The formula

= Sales[units] * Sales[price]

is inserted into the first cell of 'amount' (it happens to be J8 but that is immaterial).

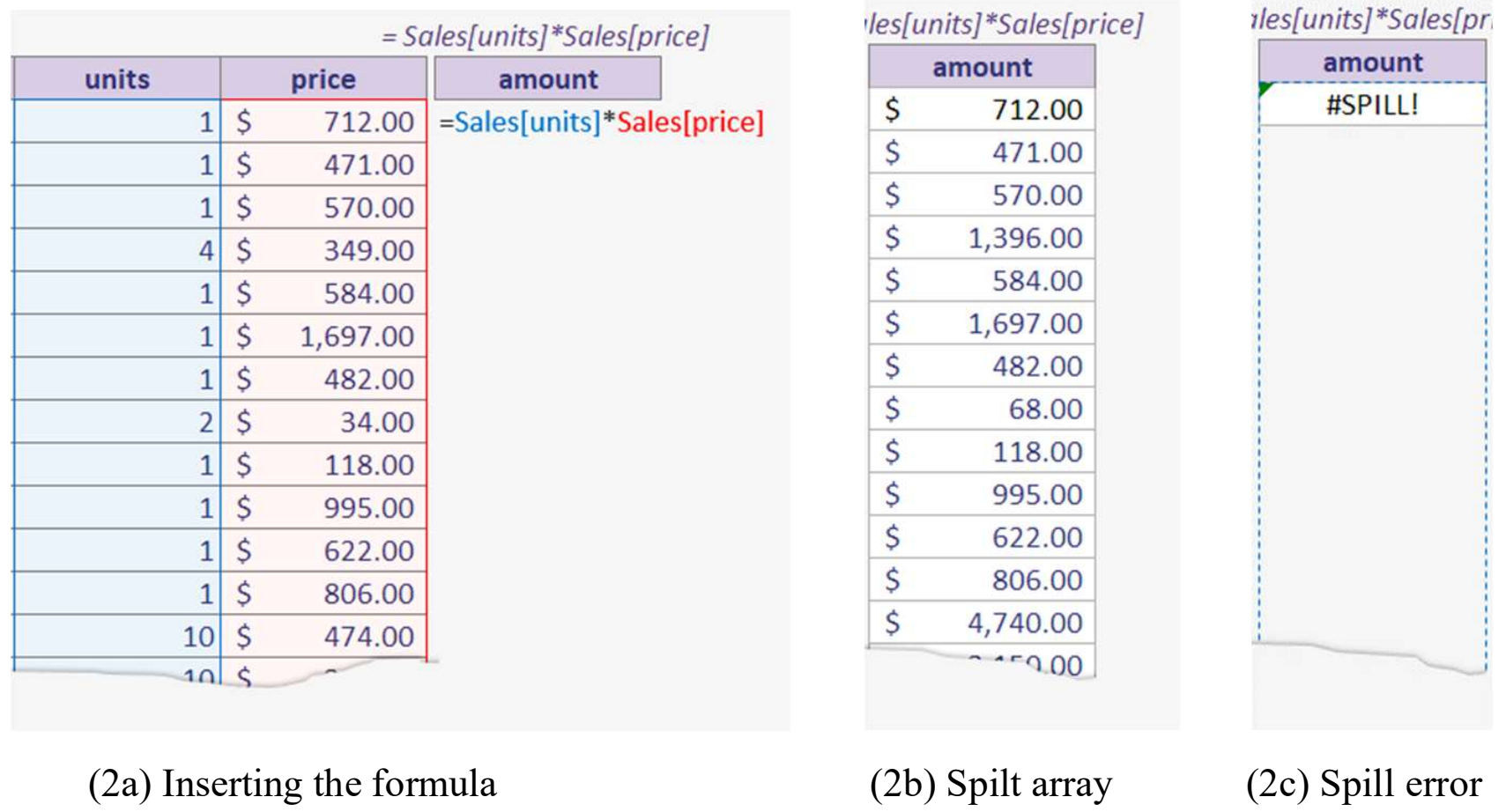

(2a) Inserting the formula (2b) Spilt array (2c) Spill error

**Figure 2 Building an Array Formula**

Whole column references to the table are chosen so that as soon as the formula is committed using the 'Enter' key the entire array of 5000 values is calculated as shown in Figure 2b.

The process of displaying the resulting array over adjacent cells is known as 'spilling'. It is not unusual for concern to be expressed at this point: 'What happens if I have important data in the way of the spilt array; will I lose it?' The data will not be lost; instead the dynamic array will simply give a #SPILL! error, as shown in Figure 2c. As soon as the offending content is deleted or moved, the array appears. Something else that is clear from the figure showing the #SPILL! error is that what appears as a full and busy worksheet can be almost entirely blank space with only a few cells populated by formulas. Next we turn to the data analysis.

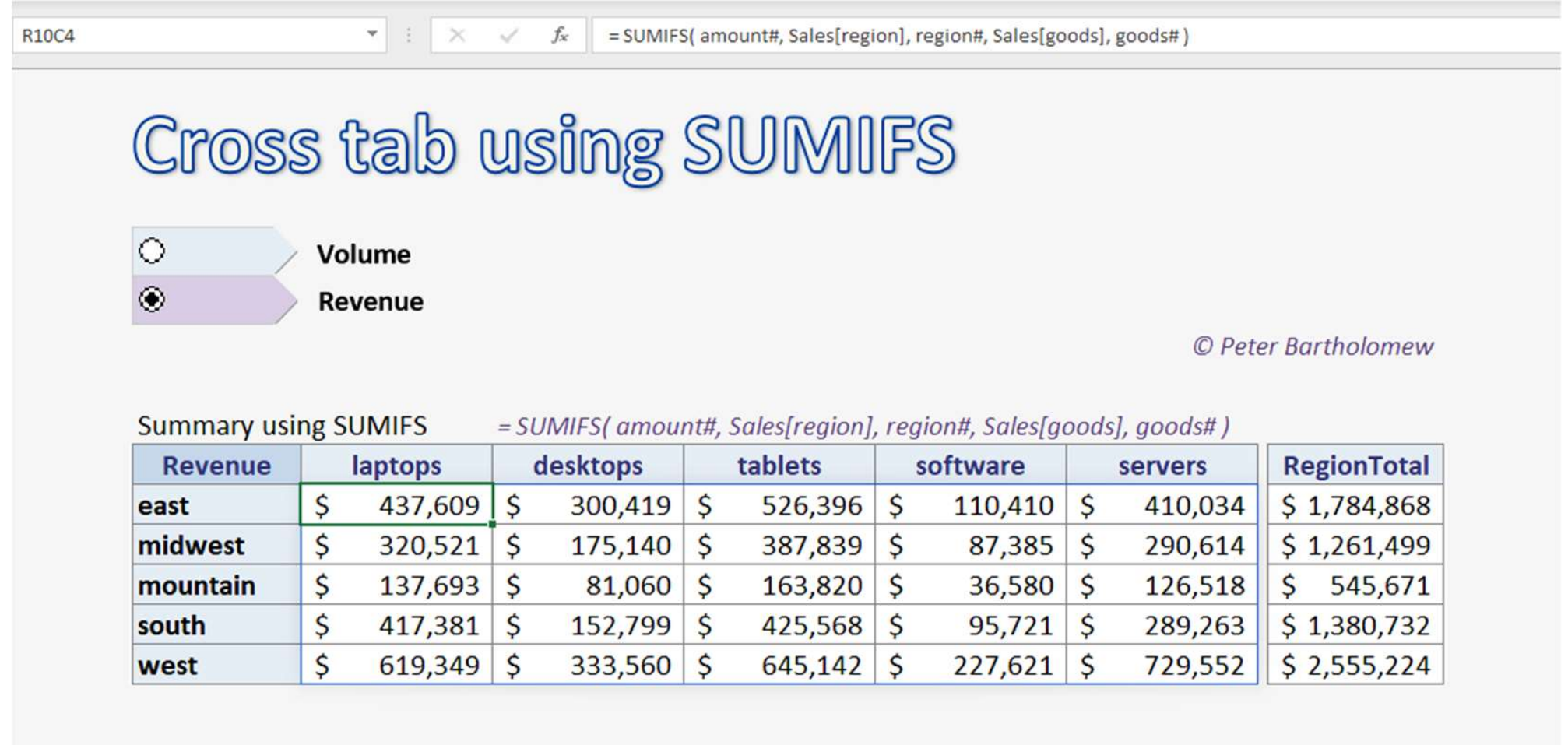

| Revenue | laptops | desktops | tablets | software | servers | RegionTotal |
|---|---|---|---|---|---|---|
| east | $ 437,609 | $ 300,419 | $ 526,396 | $ 110,410 | $ 410,034 | $ 1,784,868 |
| midwest | $ 320,521 | $ 175,140 | $ 387,839 | $ 87,385 | $ 290,614 | $ 1,261,499 |
| mountain | $ 137,693 | $ 81,060 | $ 163,820 | $ 36,580 | $ 126,518 | $ 545,671 |
| south | $ 417,381 | $ 152,799 | $ 425,568 | $ 95,721 | $ 289,263 | $ 1,380,732 |
| west | $ 619,349 | $ 333,560 | $ 645,142 | $ 227,621 | $ 729,552 | $ 2,555,224 |

Figure 3: Extract from Pivot Data by aggregation

The key to producing a simple crosstab solution that updates dynamically as the data table is extended or modified is to use the newly introduced array functions to create the row and column headers.

The first two functions turn manual data management processes into simple formulas that update automatically, thus eliminating one source of user error. The following formula yields the distinct row headers and the sorted list resizes dynamically without user intervention.

```
= SORT( UNIQUE( Sales[region]) )
```

The next formula produces the column headers as a list of distinct values but then nests this within the classic array function TRANSPOSE to give a row of headers. Now, however, *Ctrl+Shift+Enter* is not required

```
= TRANSPOSE( UNIQUE( Sales[goods] ) )
```

Now it only remains to fill in the table

```
= SUMIFS( amount#, Sales[region], region#, Sales[goods], goods# )
```

Each value is calculated by referencing an element of these two vectors as criteria and a column of the table or a range reference for the values to be summed. The Names are single cell references but the addition of the "#" symbol changes the reference from a single cell into a dynamic range reference. The formula is placed in a single cell and the 25 values are an array property of that cell. The other 24 cells of the Range are empty but they serve to display the result to the user.

Used for small to moderate sized problems the new Dynamic Array functionality works seamlessly as a means of data aggregation.

## 6 EXAMPLE – MODELLING WITH SIMILAR LINE ITEMS

This problem was developed by Levi Bailey and Vishal Rander within the context of a LinkedIn Discussion Group to test and compare the various approaches to building financial models as advocated by several leading finance modelling groups. It was presented by the present author in the 2016 EuSpRIG conference, along with a variety of non-financial models, to demonstrate the feasibility of using defined names and CSE array formulas to build models. It was concluded that such methods can provide a coherent solution strategy in which the problem is solved by a sequence of formulas resembling the steps of a programmed language. Despite that it is acknowledged that the use of CSE array formulas can be inflexible and labour-intensive.

Since that time, the Excel calculation engine has been redeveloped and Array formulas are now the default mode of calculation. Only the anchor cell contains the formula and the entire array is a property of that cell. The size of the array as displayed no longer depends upon the actions of the user; it is determined by the sizes of the referenced arrays.

In the main, this presents massive advantages since an entire array is generated by committing formula in a single cell. As will be explained, though, this problem exhibits several features that makes the application of dynamic array formulas less than routine.

One such calculation that cannot be achieved with the array formula is an accumulation in which each element of an array refers to its immediate predecessor. In mathematics this would be known as a recurrence relation and in finance it might be a 'corkscrew' or an escalation. In a reply to an email Joe McDaid confirmed that Dynamic Arrays are not suited to corkscrew style calculations because they do not support breakup – which CSE arrays do. The solution is to revert to CSE arrays for the calculation, which will mean that they are no longer dynamic, or, as is done in the example, to reformulate the problem in a manner that allows each term to be calculated without reference to the recurrence relation. There are examples of this in the spreadsheet that accompanies this paper.

The first is the time ruler. A standard way of creating the start and end of each time period is to set each start date by incrementing the end date from the previous period by one day. The end of period is then calculated as the end of month, quarter or year as required. This will generate an error as a dynamic array formula. If the time interval is constant along the ruler the simple workaround is to calculate the dates directly from the period index 'p'. The number of periods depends both on the time duration of the model and the number of months per period. The new function SEQUENCE can then be used to generate 'p'

```
=SEQUENCE(1, COLUMNS(monthlyDemand) / monthsPerPeriod, 0, 1)
```

starting at zero.

> **SEQUENCE( rows, [columns], [start], [step] )**
>
> The SEQUENCE function allows you to generate a list of sequential numbers in an array
>
> **Rows**, the number of rows in the array
>
> **Columns**, *optional*, the number of columns in the array (*defaults to* 1)
>
> **Start**, *optional*, the first number in the sequence (*defaults to* 1)
>
> **Step**, *optional*, the increment (*defaults to* 1)

The start of each period then follows immediately using

```
= 1 + EOMONTH( input.startDate, p * monthsPerPeriod )
```

This time ruler automatically resizes to match the problem description. A similar problem occurs with the product price escalation formula. Again, step by step multiplication by the escalation factor would be a normal approach. Provided the escalation rate is assumed constant, however, a simple formula will give the price at any time period

= IF( isEscalated, ( 1 + price.escalationPerPeriod# )^p, 1 ) * price.initial

In these situations, the accumulations have been replaced by the terms of an arithmetic sequence and a geometric sequence respectively. In this case the array 'p' causes the formula to spill across the sheet with minimal user input. In the latter case the initial prices and escalation factors are also arrays which specify values for each product. Therefore, the formula also spills down the sheet to give a 2D array of values.

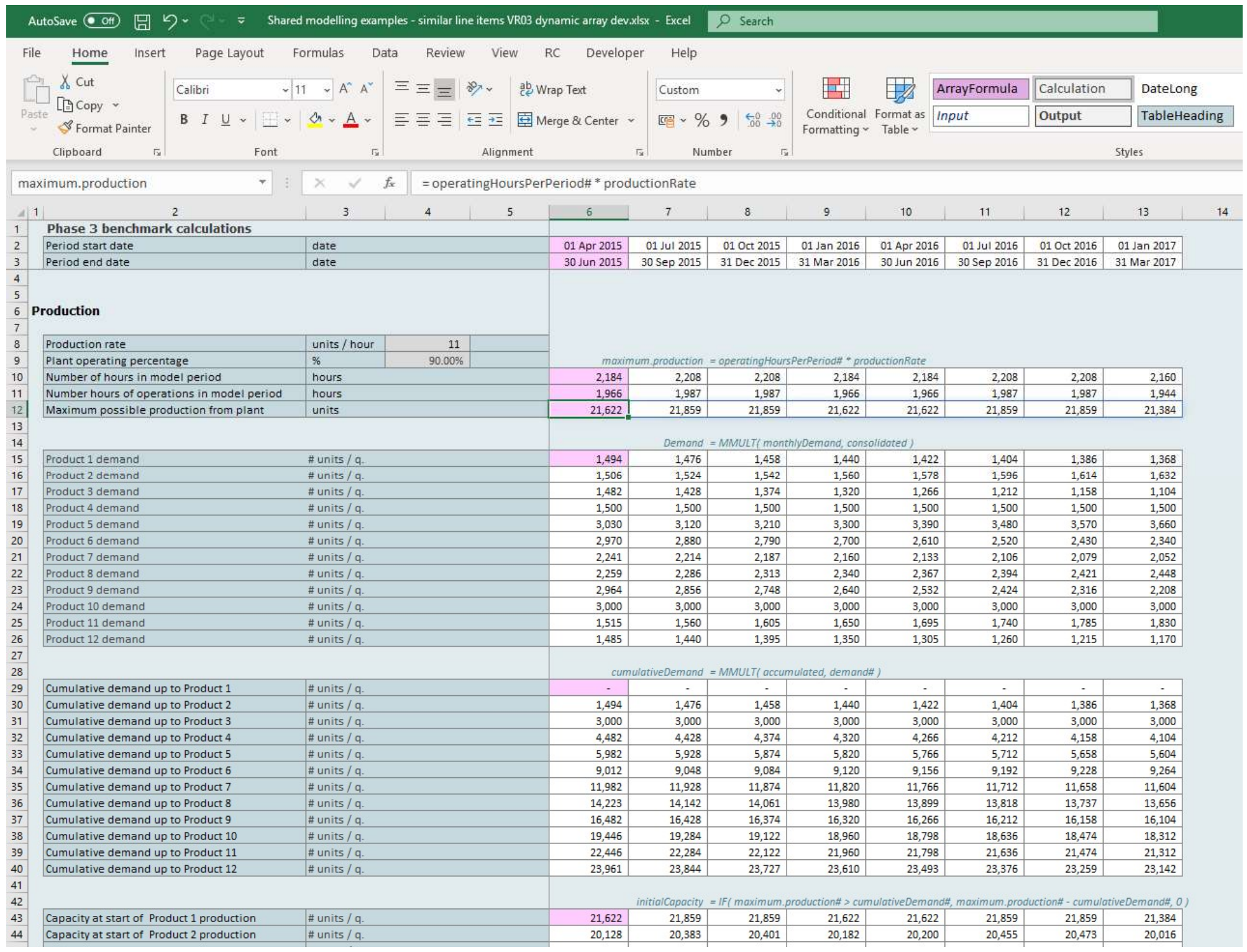

| | 2 | 3 | 4 | 6 | 7 | 8 | 9 | 10 | 11 | 12 | 13 |
|---|---|---|---|---|---|---|---|---|---|---|---|
| 1 | Phase 3 benchmark calculations | | | | | | | | | | |
| 2 | Period start date | date | | 01 Apr 2015 | 01 Jul 2015 | 01 Oct 2015 | 01 Jan 2016 | 01 Apr 2016 | 01 Jul 2016 | 01 Oct 2016 | 01 Jan 2017 |
| 3 | Period end date | date | | 30 Jun 2015 | 30 Sep 2015 | 31 Dec 2015 | 31 Mar 2016 | 30 Jun 2016 | 30 Sep 2016 | 31 Dec 2016 | 31 Mar 2017 |
| 4 | | | | | | | | | | | |
| 5 | | | | | | | | | | | |
| 6 | Production | | | | | | | | | | |
| 7 | | | | | | | | | | | |
| 8 | Production rate | units / hour | 11 | | | | | | | | |
| 9 | Plant operating percentage | % | 90.00% | maximum.production = operatingHoursPerPeriod# * productionRate | | | | | | | |
| 10 | Number of hours in model period | hours | | 2,184 | 2,208 | 2,208 | 2,184 | 2,184 | 2,208 | 2,208 | 2,160 |
| 11 | Number hours of operations in model period | hours | | 1,966 | 1,987 | 1,987 | 1,966 | 1,966 | 1,987 | 1,987 | 1,944 |
| 12 | Maximum possible production from plant | units | | 21,622 | 21,859 | 21,859 | 21,622 | 21,622 | 21,859 | 21,859 | 21,384 |
| 13 | | | | | | | | | | | |
| 14 | | | | Demand = MMULT( monthlyDemand, consolidated ) | | | | | | | |
| 15 | Product 1 demand | # units / q. | | 1,494 | 1,476 | 1,458 | 1,440 | 1,422 | 1,404 | 1,386 | 1,368 |
| 16 | Product 2 demand | # units / q. | | 1,506 | 1,524 | 1,542 | 1,560 | 1,578 | 1,596 | 1,614 | 1,632 |
| 17 | Product 3 demand | # units / q. | | 1,482 | 1,428 | 1,374 | 1,320 | 1,266 | 1,212 | 1,158 | 1,104 |
| 18 | Product 4 demand | # units / q. | | 1,500 | 1,500 | 1,500 | 1,500 | 1,500 | 1,500 | 1,500 | 1,500 |
| 19 | Product 5 demand | # units / q. | | 3,030 | 3,120 | 3,210 | 3,300 | 3,390 | 3,480 | 3,570 | 3,660 |
| 20 | Product 6 demand | # units / q. | | 2,970 | 2,880 | 2,790 | 2,700 | 2,610 | 2,520 | 2,430 | 2,340 |
| 21 | Product 7 demand | # units / q. | | 2,241 | 2,214 | 2,187 | 2,160 | 2,133 | 2,106 | 2,079 | 2,052 |
| 22 | Product 8 demand | # units / q. | | 2,259 | 2,286 | 2,313 | 2,340 | 2,367 | 2,394 | 2,421 | 2,448 |
| 23 | Product 9 demand | # units / q. | | 2,964 | 2,856 | 2,748 | 2,640 | 2,532 | 2,424 | 2,316 | 2,208 |
| 24 | Product 10 demand | # units / q. | | 3,000 | 3,000 | 3,000 | 3,000 | 3,000 | 3,000 | 3,000 | 3,000 |
| 25 | Product 11 demand | # units / q. | | 1,515 | 1,560 | 1,605 | 1,650 | 1,695 | 1,740 | 1,785 | 1,830 |
| 26 | Product 12 demand | # units / q. | | 1,485 | 1,440 | 1,395 | 1,350 | 1,305 | 1,260 | 1,215 | 1,170 |
| 27 | | | | | | | | | | | |
| 28 | | | | cumulativeDemand = MMULT( accumulated, demand# ) | | | | | | | |
| 29 | Cumulative demand up to Product 1 | # units / q. | | - | - | - | - | - | - | - | - |
| 30 | Cumulative demand up to Product 2 | # units / q. | | 1,494 | 1,476 | 1,458 | 1,440 | 1,422 | 1,404 | 1,386 | 1,368 |
| 31 | Cumulative demand up to Product 3 | # units / q. | | 3,000 | 3,000 | 3,000 | 3,000 | 3,000 | 3,000 | 3,000 | 3,000 |
| 32 | Cumulative demand up to Product 4 | # units / q. | | 4,482 | 4,428 | 4,374 | 4,320 | 4,266 | 4,212 | 4,158 | 4,104 |
| 33 | Cumulative demand up to Product 5 | # units / q. | | 5,982 | 5,928 | 5,874 | 5,820 | 5,766 | 5,712 | 5,658 | 5,604 |
| 34 | Cumulative demand up to Product 6 | # units / q. | | 9,012 | 9,048 | 9,084 | 9,120 | 9,156 | 9,192 | 9,228 | 9,264 |
| 35 | Cumulative demand up to Product 7 | # units / q. | | 11,982 | 11,928 | 11,874 | 11,820 | 11,766 | 11,712 | 11,658 | 11,604 |
| 36 | Cumulative demand up to Product 8 | # units / q. | | 14,223 | 14,142 | 14,061 | 13,980 | 13,899 | 13,818 | 13,737 | 13,656 |
| 37 | Cumulative demand up to Product 9 | # units / q. | | 16,482 | 16,428 | 16,374 | 16,320 | 16,266 | 16,212 | 16,158 | 16,104 |
| 38 | Cumulative demand up to Product 10 | # units / q. | | 19,446 | 19,284 | 19,122 | 18,960 | 18,798 | 18,636 | 18,474 | 18,312 |
| 39 | Cumulative demand up to Product 11 | # units / q. | | 22,446 | 22,284 | 22,122 | 21,960 | 21,798 | 21,636 | 21,474 | 21,312 |
| 40 | Cumulative demand up to Product 12 | # units / q. | | 23,961 | 23,844 | 23,727 | 23,610 | 23,493 | 23,376 | 23,259 | 23,142 |
| 41 | | | | | | | | | | | |
| 42 | | | | initialCapacity = IF( maximum.production# > cumulativeDemand#, maximum.production# - cumulativeDemand#, 0 ) | | | | | | | |
| 43 | Capacity at start of Product 1 production | # units / q. | | 21,622 | 21,859 | 21,859 | 21,622 | 21,622 | 21,859 | 21,859 | 21,384 |
| 44 | Capacity at start of Product 2 production | # units / q. | | 20,128 | 20,383 | 20,401 | 20,182 | 20,200 | 20,455 | 20,473 | 20,016 |

**Figure 4 Extract from "Shared Modelling Example with Similar Line Items**

A more severe accumulation problem arises when one tries to accumulate the demand over the diverse products to compare the result with the total production capacity for each period. Previously a CSE array formula was used with the referenced array being offset one row from the result. Because this required array breakup, an alternative solution was sought using matrix multiplication. This used a lower triangular matrix in which the values below the leading diagonal were all ones. Multiplied into the demand array, this first gives the top row, then the sum of the top and second rows and so on. The matrix itself is build using the SEQUENCE array function within the named formula 'accumulate' that refers to

= SIGN( SEQUENCE(1,12) < SEQUENCE(12,1) )

The cumulative demand array is then given by

= MMULT( accumulate, demand# )

where 'demand#' is the notation for a spilt array anchored at a cell named 'demand'. MMULT implements the mathematical operation of matrix multiplication.

From there the residual capacity is

```
= IF( maximum.production# > cumulativeDemand#,
      maximum.production# – cumulativeDemand#, 0 )
```

and the volume produced for each product is

```
= IF( demand<initialCapacity#, demand#, initialCapacity# )
```

A striking feature of the extract from the workbook shown in Figure 4 are the magenta cells. These are produced by a conditional format based upon the formula

```
=ISFORMULA(RC)
```

The surprise for anyone versed in traditional methods is how few cells contain formulas.

The second problem to be overcome when building solutions using array formulas is to aggregate a 2D array either row-by-row or column-by-column. Pretty much all the aggregation operators such as SUM, PRODUCT, SMALL, MIN, LARGE, MAX, AND, OR produce a single result from a 2D array yet it is perfectly reasonable to want to apply such formulas to each row or each column. The expectation is that the user will use relative referencing to select rows or columns that they wish to aggregate manually. This, however, means that the column of row sums, for example, does not expand dynamically as additional rows are added to the data. The one exception to this limitation is MMULT which will perform batches of weighted sums either row-wise or column-wise. The result obtained from MMULT will expand dynamically, provided care is taken to avoid non-numeric fields and to ensure the internal dimensions always match exactly.

In the present case, the total product revenues are calculation using

```
= MMULT( TRANSPOSE(active), product.revenue# )
```

where 'active' is a Boolean array that is 1 where data has been provided for the specific product and 0 if it has been omitted.

This solution is effective but is unnecessarily complicated in terms of its mathematics. In my mind it would be better to have a function that indicates that the aggregation should be carried out column by column but returned as a row array of results. Suggestions might include

```
= SUM( BYCOLUMN( product.revenue# ) )
```

or, if a more compact notation were required

```
= SUM( |product.revenue# )
```

where the 'pipe' character here indicates a column.

## 7 ELEMENTS OF RISK

As yet there is virtually no experience of applying the methods outlined within the Excel community, so no evidence exists to back up statements concerning risk. However, using the Panko (2010) taxonomy of spreadsheet errors, Figure 5, the following might be considered to provide reasonable expectations.

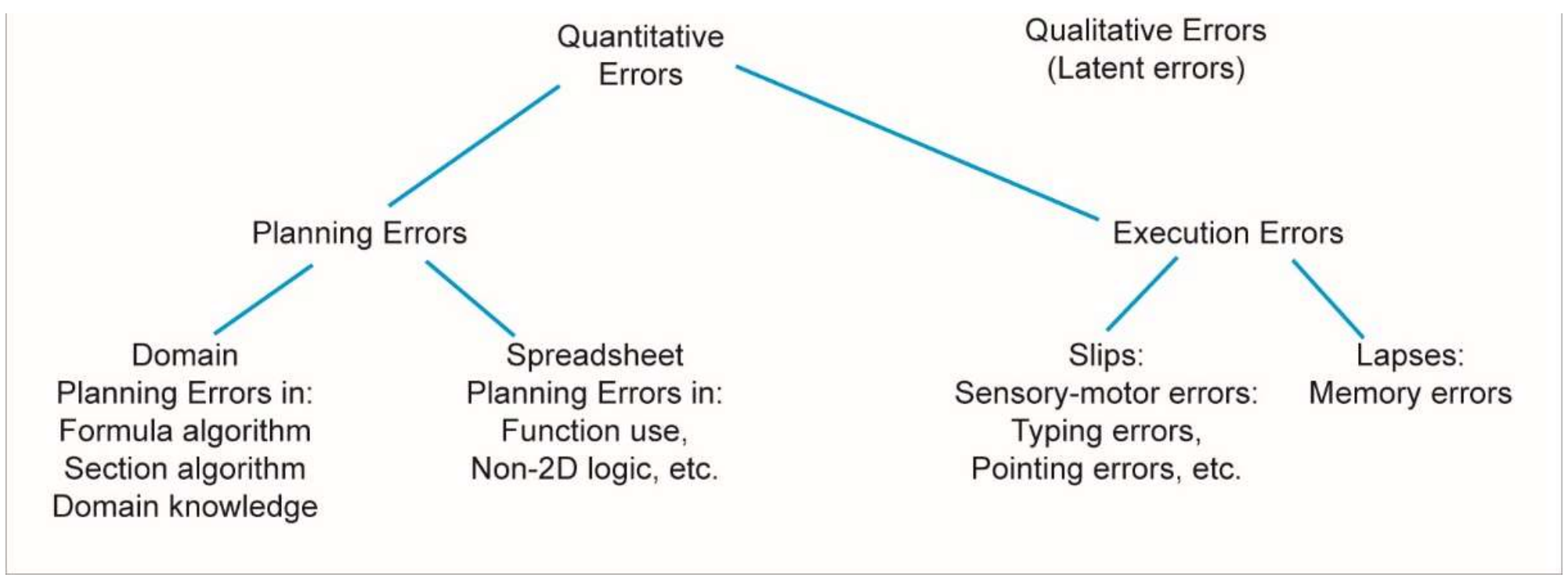

**Figure 5 Taxonomy of Spreadsheet Errors**

Firstly, considering execution errors, the fact that dynamic arrays require an order of magnitude fewer formulas must reduce the opportunity for error. Typing errors and pointing errors, in particular, may be expected to reduce.

Some typing errors manifest themselves as consistency errors in which a region of formulae that may be intended to be the same have cells that differ from the surrounding region. Such inconsistencies can also arise from ill-conceived attempts to correct a specific instance of the formula or the data it references. Dynamic arrays eliminate such errors; consistency errors are simply not possible when only one instance of a formula is present.

Sizing errors, whereby only part of a range is generated, or subsequently referenced within an aggregation formula, may also be expected to reduce since these steps no longer require user interaction.

The consistent use of Names may also be expected to have a role in reducing the effect of typing errors in that wrongly typed names are more likely the generate #NAME! errors than erroneous references. In particular, erroneous references to 'white space' that has not been assigned to the model are eliminated.

Turning now to planning errors: it is still quite possible to introduce errors in the application of Domain knowledge. That said, the consistent use of meaningful names creates formulae that are semantically meaningful. This, in turn, may be expected to reduce the likelihood of such errors passing unnoticed.

A potential disadvantage of programming Excel through the use of named arrays is that the more abstract level of thinking required is alien to most practitioners. It is thought likely that planning errors associated with inadequate Excel knowledge will actually increase.

## 8 DISCUSSION AND CONCLUSIONS

The introduction of modern dynamic arrays could change the nature of spreadsheet development utterly. No longer need it be a process of manual manipulation of large blocks of data by reference to the rows and columns they occupy. As has been shown, dynamic formulas typically reside in a single cell and only appear to fill large regions of the worksheet for presentational purposes.

There remains scant need for the concept of relative referencing or for the default A1 notation introduced in the original spreadsheet concept. Some additional work is still required on behalf of the developer to name each formula, now occupying a single cell, but it requires little more effort than simply labelling the cell, as would be normal good practice. Referencing raw data, whether input directly to Excel or read in *via* Power Query, does require the use of multi-cell ranges but there, assuming Tables are used extensively, naming is largely taken care of by structured referencing.

The advantage of fully dynamic models is that they require less manual intervention to keep them updated or, alternatively, to modify them to accept new data sets. This reduces risk and the attendant errors. The advantage of consistent naming conventions is that the formulas are semantically meaningful and can be readily checked for correctness in terms of the both the Excel logic and, more importantly, the domain knowledge that links the proposed solution to the business problem. Consistency errors disappear because the formula is only written within a single cell. Array sizing errors in which the final terms of an array are accidentally omitted are also largely eliminated because it is the code that sizes the array and not the user. With DA, array formulas become the path of least resistance rather than some obscure method of last resort.

What is lost is the ability to identify relationships between individual cells using precedent trees and to check their correctness using a pocket calculator. I believe that is small loss, to check a representative sample sufficiently to estimate the probability of the workbook being correct would be a mind-numbingly tedious and repetitive task. Moreover, it is addressing the wrong problem. It is rare that Excel produces an incorrect result for a correctly formulated formula; it is far more likely that the formula is incorrectly implemented or does not capture the business logic.

So, having argued that modern Excel is capable of revolutionising spreadsheet practice, yielding something far more akin to a rigorous programming approach that the *ad-hoc* style of end user programming, the question remains "Will such a revolution actually come to pass?"

By providing backward compatibility one ensures that it is still possible to develop workbooks in the old way, that of replicating single-cell formulas. Users can just continue the way they always have, oblivious to all new functionality as they have, by and large, done in the case of Tables, Pivot tables, Power Query *etc*. If we assume that most users, who are not aware that Excel involves programming and do not see its relevance, will not move beyond the turn of the century, how should developers respond? I would argue that developers should no longer accept the limitations of their client's knowledge as constraints on the techniques they employ. If a client wants an *ad-hoc* programming style they are at liberty to do it themselves. My experience is that, once a structured Excel application has been delivered to the client, it will over time acquire unstructured additions as a result of client interaction. Sooner or later it will go wrong and at that point there exists an opportunity for the developer to refactor these changes to provide new functionality.

Ultimately a well-constructed 'app' built on Excel as a platform provides a far more controlled risk environment than a loosely structured worksheet that everyone feels free to modify.